\documentclass[1p]{elsarticle}
 \pdfoutput=1

\usepackage{latexsym}
\usepackage{graphics}
\usepackage{amssymb}
\usepackage{subfigure}
\usepackage{textcomp}
\usepackage{gensymb}
\usepackage[english]{babel}
\usepackage[T1]{fontenc}
\usepackage{setspace}
\usepackage{color}
\usepackage{siunitx}
\usepackage{isotope}
\usepackage{here}
\usepackage{afterpage}

\journal{Nucl. Instr. and Meth. A}

\begin{document}

\begin{frontmatter}

\title{The MCUCN simulation code for ultracold neutron physics}

\author{G. Zsigmond\corref{mycorrespondingauthor}}


\cortext[mycorrespondingauthor]{Corresponding author}
\ead{geza.zsigmond@psi.ch}

\address[mymainaddress]{Laboratory for Particle Physics, Paul Scherrer Institut, CH-5232 Villigen-PSI, Switzerland}

\begin{abstract}
Ultracold neutrons (UCN) have very low kinetic energies 0-300 neV, thereby can be stored in specific material or magnetic confinements for many hundreds of seconds. This makes them a very useful tool in probing fundamental symmetries of nature (for instance charge-parity violation by neutron electric dipole moment experiments) and contributing important parameters for the Big Bang nucleosynthesis (neutron lifetime measurements). Improved precision experiments are in construction at new and planned UCN sources around the world. MC simulations play an important role in the optimization of such systems with a large number of parameters, but also in the estimation of systematic effects, in benchmarking of analysis codes, or as part of the analysis. The MCUCN code written at PSI has been extensively used for the optimization of the UCN source optics and in the optimization and analysis of (test) experiments within the nEDM project based at PSI. In this paper we present the main features of MCUCN and interesting benchmark and application examples.
\end{abstract}


\end{frontmatter}


\section{Introduction}
\label{label-introduction}

Ultracold neutrons (UCN) have kinetic energies below about 300\,neV and experience total reflection by specific material walls, e.g. nickel, at any angle of incidence. The condition for this is that the optical potential of the wall material originating from coherent nuclear scattering \cite{Fermi1936,Fermi1946} must be larger than the kinetic energy of the neutrons. This very low energy range is also comparable to the gravitational potential energy of the neutron: a 102.5\,neV UCN flying vertically can only reach a height of one meter. A third interaction, the force exerted on UCN by a magnetic field gradient, can be used as well to confine them in a storage volume. A magnetic field change of 2 T along the UCN path causes 120.6\,neV change in its kinetic energy. The magnetic interaction coupled to the particle's magnetic moment (the Larmor frequency is 29.2\,Hz in a 1\,$\mu$T field) makes neutrons at the same time to sensitive magnetometers. 

These properties allowing to store and manipulate UCN make them to excellent probes in low-energy particle physics, since longer storage times imply increased observation times and thus, under stable conditions, also higher precision. The most important UCN experiments probe fundamental symmetries of nature (e.g. charge-parity violation by neutron electric dipole moment experiments \cite{Pendlebury2015,SerebrovEDM2015,Kasprzak2016}) or yield important precise parameters for the Big Bang nucleosynthesis (neutron lifetime measurements \cite{Morris2017,Serebrov2008,Leung2016,ntaubeam2013,MATERNE2009}). In these areas new, improved precision experiments are in construction at all operating UCN sources around the world \cite{Bison2017}. 

In UCN physics experiments Monte Carlo (MC) simulation methods are very effective in the characterization and optimization of many-parameter systems, and when a detailed calculation of a geometry containing several sub-volumes is relevant. Exact analytic modeling is not feasible in many cases $-$ for example because of gravity effects on UCN or because of UCN viewing several surface qualities at the same time $-$ or it would involve multiply-coupled differential equations with a large number of parameters. Such complex analytic models would also need as final tests additional MC benchmarking.

The MCUCN code has been developed first in collaboration with the UCN physics group at PSI as input for the phase space transformer project at SINQ \cite{Mayer2009}, and subsequently further improved supporting the projects of this group, e.g. \cite{LAUSS201498,Lauss2012,Afach2015PRD,ATCHISON2007647}. 

MCUCN has been extensively used in the optimization of the PSI UCN source, beamlines and experiments in order to maximize UCN density or transmission \cite{LAUSS201498,Lauss2012,Blau2017SourceBeamlines,Blau2017Pingpong}. MCUCN has already been a useful tool in estimating systematic effects in the neutron electric dipole moment (nEDM) experiment \cite{Pendlebury2015,Afach2015PRL,Afach2015PRD}, for checking data analysis software by toy data, or as part of the data analysis (providing e.g. collision rates \cite{Bondar2017}). Future MC simulations will also be deployed for helping the planning of time-efficient experiments e.g. by improving the online analysis with MC data. 

Several alternative simulation codes are available as STARucn \cite{STARucn}, PENTrack \cite{Schreyer2017123}, and GEANT4UCN \cite{G4UCN}. This allows for comparison of independent calculations, very important for increasing confidence in using these codes and in intricate numerical estimations. Inter-comparison calculations involving MCUCN were very successful and will be reported in a future publication.

The structure of the paper is the following: After a description of the code, and most important benchmarks with analytic models testing the simulation accuracy, we will briefly present several application examples using MCUCN.

\section{The MCUCN code}
\label{label-description}

MCUCN is a pure C++ code and thus platform independent. For the sake of transparency for non-expert programmers as well, the default version is based on functions instead of classes $-$ the latter may be added by C++ expert users when implementing specific features. Emphasis was put on making the MCUCN clearly organized and thus easily accessible for modifications and extensions. Input parameters including geometry data were concentrated into one header file. However, usage of command options for frequently changed parameters is straightforward. 

The default version of MCUCN uses standard libraries and compiles within seconds without the need of any software pre-installation $-$ an important advantage when using computing grids. The executable can thus be readily forwarded for grid-jobs running independently with separate batches of trajectories. We have used extensively the PL-Grid \cite{PLGrid} in our applications. 

The default random generator for sampling probability distributions is RAN3 \cite{Press2007}, and is replaceable by the user. 

Information on the computing speed, which very much depends on the concrete model configuration, will be given at several example calculations.

Visualization of the MCUCN output data and geometry is left to the user who can chose her or his favorite tool.

\subsection{Geometry configuration}
\label{label-Geometry-configuration}

A storage volume or UCN guiding geometry can be modeled using a combination of an arbitrary large number of reflecting or transmitting second order surface sections (planes, cylinders, cones, etc.). Toroidal parts, like bent guides, can be easily built recursively via loops from smaller second order sections without strongly increasing computing time. Time dependent shutters were also defined. Triangulated mesh exports from CAD programs in form of STL ASCII files can also be used as input for MCUCN, however, an increase in computing time has to be taken into consideration.
 
It is straightforward to configure any geometry by translation and rotation of the surface elements. In MCUCN several surfaces can be grouped into a common local coordinate system, and moved and/or rotated together thus enabling to work with an easier manageable subsystem.

\subsection{Trajectories calculation}
\label{label-Trajectories-calculation}

UCN are represented in MCUCN by time, position, velocity and spin vector coordinates. These coordinates can be written out at user-defined 'snapshot' times or detected by a virtual monitor surface, for example at a beamline exit, or after a detector window. As an option, reflection points on the surfaces and trajectory point coordinates can be generated, aiming to test the interaction of UCN with the surfaces of the given confinement, by visualizing them with an independent graphics tool.

The random initial coordinates are generated according to a predefined (analytic or numerical) energy distribution of UCN exiting a virtual flat (default) surface. There is an option of adding a vertical or scattered energy boost caused by the optical potential of the UCN converter material, at PSI the solid deuterium \cite{Altarev2008PRL}. 

Gravity playing an important role in UCN kinetics had to be implemented as well. In MCUCN, in contrast to GEANT4UCN, the ballistic trajectories are analytically calculated between successive reflection or transmission points, this making MCUCN calculations very fast. 

The successive reflection points are calculated by using the classical equation of motion in the presence of gravity, and substituting this into the second order equation for each surface defined in the geometry configuration. By solving the resulting quartic equation \cite{Roots3And4} in time for every surface and by checking that the interaction point is within the surface boundaries one can find the next surface hit by the UCN. This is done by choosing the surface corresponding to the smallest flight time. Using this flight time value, the final spacial and velocity coordinates of a ballistic trajectory are obtained. These will be subsequently used for calculating the  interaction with the surface hit by the UCN.

Motion in a strong magnetic field deflecting the UCN trajectories has not yet been generally implemented in MCUCN, but a local module can be called calculating the UCN path in a cylindrical volume representing a strong local field like a magnetic valve. This module introduces small trajectory steps and thus strongly decreases  the computing speed. Special cases, for example a superconducting magnet polarizer, or a magnetized foil were approximated in our calculations by a spin-dependent rectangular potential energy barrier or well, depending on the spin state, by adding an energy boost or deceleration of 60.3\,neV/T. Comparisons to measured transmissions when ramping our superconducting polarizer magnet \cite{Blau2017SourceBeamlines} made us confident that this approach is also valid.

\subsection{Interaction with walls}
\label{label-Interaction}

The UCN interaction properties on the surfaces are implemented according to textbook UCN physics \cite{Golub1991}. It is straightforward to add new features by numerical input or by modifying the current default analytic formulas. 

The standard input parameters related to UCN-surface interactions are: 

\begin{itemize}
\item Volume averaged coherent nucleon-neutron potential (aka Fermi or optical potential) 

\begin{equation}
    V_F = \frac{2 \pi \hbar^2}{m} N b_\text{coh} ,
\end{equation}

where $N$ is the density of nuclei, $b_\text{coh}$ the coherent scattering length, and $m$ the mass of the neutron.
\item Material loss constant $\eta = W_\text{loss} / V_\text{F} $, where 

\begin{equation}
W_{loss} =  \frac{\hbar}{2} N \sigma_{loss} v
\end{equation}

is the complex potential associated with the absorption and up-scattering cross section $\sigma_\text{loss}$ which is inversely proportional to the particle velocity in the medium $v$. For materials with mixed scattering species, $b_\text{coh}$ and $\sigma_\text{loss}$ have to be replaced by their volume averages.  The energy and direction dependent neutron reflectivity is calculated from the complex modulus of the reflected wavefunction amplitude as given in eq. (2.71) of \cite{Golub1991}

\begin{equation}
\label{eq-reflectivity}
		\left| R \right| ^2 = \frac
												{E_\perp - \sqrt{E_\perp} \sqrt{2\alpha-2\left(V_\text{F}-E_\perp \right)} + \alpha}
												{E_\perp + \sqrt{E_\perp} \sqrt{2\alpha-2\left(V_\text{F}-E_\perp \right)} + \alpha},
\end{equation} 

where $E_\perp$ is the perpendicular part of the kinetic energy, and 

\begin{equation}
\alpha = \sqrt{\left(V_\text{F}-E_\perp \right)^2+W^2}. 
\end{equation} 

Eq. (\ref{eq-reflectivity}) is valid for all $E_\perp$ values both below and above the optical potential $V_\text{F}$.

\item Gap loss probability per bounce representing a uniformly distributed loss probability along the surface. However, individual gaps can also be implemented by the geometry settings.

\item For non-specular reflections we use per default a weighting parameter $p_\text{diff}$. With this we define the fraction of ideally diffuse reflections described by the Lambert model discussed in detail in subsection 4.4.5 in \cite{Golub1991}. For the UCN fraction which is reflected diffusely, the probability distribution of $\text{cos}\theta$ (the projection of the reflected direction vector onto the surface normal) is linear. This approach is applicable if macroscopic surface irregularities and cavities dominate over microscopic roughness which leads to a practically instant memory-loss regarding the previous flight direction.  However, a second option, a detailed micro-roughness model, as formulated by \cite{Steyerl1972}, can also be chosen. This latter model describes interference of reflected waves if the roughness is comparable to the neutron wavelength.  This roughness option has been benchmarked supporting previous work published in \cite{Atchison2010}. It is rarely used because of extensive computing time and because it is only applicable for highly polished surfaces (roughness everywhere below 5\,nm excluding any local high roughness).

\item Attenuation parameter for foil transmission, which was defined as the product of the macroscopic attenuation constant (cm$^{-1}$) and the foil thickness, for 1\,m/s velocity calculated inside the transmitting medium. Partial reflection is also considered at the exiting surface of a foil according to standard formulas for quantum reflection and transmission \cite{Steyerl1971}, while neglecting loss effects and multiple reflections between the two faces of the foil:

\begin{equation}
\left| T \right| ^2 = \frac
												{4v_\text{vac}v_\text{foil}}
												{\left( v_\text{vac}+v_\text{foil} \right)^2},
\end{equation}

\begin{equation}
\left| R \right| ^2 = 1 - \left| T \right| ^2 
\end{equation}

where $v_\text{vac} = \sqrt{2 E_\perp / m} $ and $v_\text{foil} = \sqrt{2 (E_\perp - V_\text{F} )/ m)}$ are the velocities in vacuum and within the foil, respectively. Here we also assume that the foil has macroscopic thickness and thus no interference effects occur, i.e. the foil thickness is much larger than the wavelength of UCN. Optionally, the flight direction after exiting the foil can be scattered according to a Gaussian distribution.

\item Spin-flip probability per reflection.
\end{itemize}

\subsection{Spin precession}
\label{label-Spin-precession}

Low magnetic fields which are not deflecting ($ E_\text{kin} \gg \left| \mu B \right| $) were implemented using the Bloch equations $\partial _t \textbf{S} = - \gamma \textbf{B}  \times \textbf{S}$ for calculating the spin precession along the ballistic trajectories. We solve these coupled differential equations using the Runge-Kutta-Fehlberg method \cite{Press2007}. In this algorithm the 4$^{th}$ and 5$^{th}$ Runge-Kutta orders are calculated simultaneously, their difference being the truncation error which is then compared to the required precision (an input parameter, the allowed maximal absolute error per step for each spin vector component) to adapt the step-size in time. In our example simulations the Runge-Kutta-Fehlberg method proved to be several times faster than the classical Runge-Kutta-4 method. 

The spin handling algorithm by introducing small time-steps added to the fast ballistic calculations increases the total computing time. We will give computing speed examples in the test calculations presented below.

The input magnetic field has to be defined analytically, and can be an arbitrary function of time, space and velocity coordinates. Velocity dependence is realized e.g. in the relativistic motional field effect $\textbf{B}_v = (\textbf{E}\times \textbf{v})/c^2$, important for nEDM experiment systematics. 

Spin precession in any kind of inhomogeneous magnetic field can be calculated by reading in coefficients of multipole orders using a predefined harmonic decomposition \cite{Pignol2012}. The field parametrization can be obtained by a separate code by fitting measured magnetic field data.

\section{Benchmark examples}
\label{label-benchmarks}

We tested MCUCN against analytical calculations to demonstrate that the UCN physics features listed in the previous section, were correctly implemented. Further comprehensive experimental comparisons were performed which will be published in dedicated articles \cite{Blau2017SourceBeamlines,Blau2017Pingpong}. Several first experimental benchmarks were already presented in \cite{Bodek2011}.

In our analytic benchmark tests we considered a cylinder chamber ($H = $120\,mm, $r = $235\,mm) with its axis aligned to gravity, a geometry used in the nEDM experiment \cite{Pendlebury2015}. The UCN were generated at the bottom of the chamber. The generated initial position distribution was set homogeneous. The angular distribution of the initial velocity vector was set linear in the cosine of the angle to the normal vector of the surface, reproducing the case of mechanical equilibrium (isotropic gas).

First we present tests of the UCN optics and subsequently tests of the spin handling algorithm. For details on the physics connected to these examples we refer the reader to related publications cited in the subsections.

\subsection{Neutron optics tests}
\label{label-benchmarks-optics}

UCN trajectories between two reflection points are calculated in MCUCN based on the classical equation of motion in a gravitational field. Checking whether energy conservation was respected after several ten thousand reflections serves as a first test to analyze the precision of implementation of the equation of motion, and of the coordinates calculation of reflection points on the various surfaces. After 40'000\,s storage in the cylinder chamber we obtained a shift in the total energy of $2.17\cdot 10^{-11}$\,neV with a standard deviation (SD) of $2.58\cdot 10^{-11}$\,neV. In a complex geometry as planned for the future n2EDM experiment at PSI, which will be mentioned later on in the applications section, we obtained after 300\,s (an exaggerated time for filling the chambers) an energy shift of $12.3\cdot 10^{-4}$\,neV with a corresponding SD of $1.20\cdot 10^{-4}$\,neV. 

Since UCN may be reflected many thousand times, it is also important to examine whether the finite algorithm precision (due to rounding errors) introduces 'leaks' where UCN can be lost. For such a test we switched of all physical losses for UCN and checked how long they can be stored in a closed volume. Such a 'leak test' in the cylinder chamber resulted in a storage time constant > 80'000\,s. For the much more complicated n2EDM geometry we got 40'000\,s as a lower limit.

In the examples below we intended to verify the kinetics calculations first without diffuse scattering on the walls. Next we checked how diffusely scattering walls influence the deviations from the analytical approximations. Finally, the implementation of wall losses was tested by comparing to an analytic approach involving surfaces of different parameters.

\subsubsection{Specular reflections}
\label{label-Specular-reflections}

In this test we calculated the average vertical coordinate, i.e. the center of mass of the UCN in the nEDM storage chamber, with a vertical part of the kinetic energy $E_\perp=m v^2_\perp/2$ fixed at the bottom. The horizontal velocity component was calculated from the vertical one using a random angle according to the cosine distribution. 

This is a  check of the ballistics treatment with only specular reflections when mechanical equilibrium is not granted. In case of no diffuse reflections the vertical and horizontal motion can be decoupled for each trajectory.  Thus looking at a $E_\perp$ dependence rather than the total kinetic energy is more reasonable, while the shape of the trajectories remains general.

This case can easily be calculated analytically without any assumptions, and then compared to MC simulations. 

In the simulations we switched off all losses to obtain enough statistics in a short time. The UCN were generated homogeneously over a time period (0 - 90\,s) and the average of the vertical coordinate at a later time moment (100\,s) was calculated. 

The center of mass was calculated analytically based on the condition that in mechanical equilibrium the probability to find a UCN at a height $z$ is constant in time, i.e. $P(t)dt=\text{const} \cdot dt$. Transforming the time variable into vertical coordinate $z$ it holds $P'(z,E_{\perp})dz=const \cdot dt$, because a probability element is independent of our choice of the variable. Using the equation of motion for a particle in gravitational field in the form $z(t,E_\perp)= \sqrt{\frac{2 E_\perp}{m}} t-\frac{g t^2}{2}$, where $m$ is the mass of the particle, we obtain the distribution function

\begin{equation}
    P'(z,E_{\perp}) = \left\{ 
		\begin{array}{cl}
		(1-z/h)^{-1/2}, & \text{if}\ z < \text{min}(h,H)	\\
		0, & \text{if}\ z \ge \text{min}(h,H)
		\end{array}	\right.
\end{equation}

where $h=E_\perp/mg$.

Calculating the average of the vertical coordinate we can express the center of mass of UCN:

\begin{equation}
    z_{CM}(h) = \left\{ 
		\begin{array}{cl}
		\frac{h}{3} \cdot \frac{(1-H/h)^{3/2}-3(1-H/h)^{1/2}+2}{1-(1-H/h)^{1/2}}, & \text{if}\ h>H	\\
		\frac{2 h}{3}, & \text{if}\ h \le H
		\end{array}	\right.
\end{equation}

\begin{figure*}[htb]
\begin{center}
\resizebox{0.90\textwidth}{!}{
\includegraphics{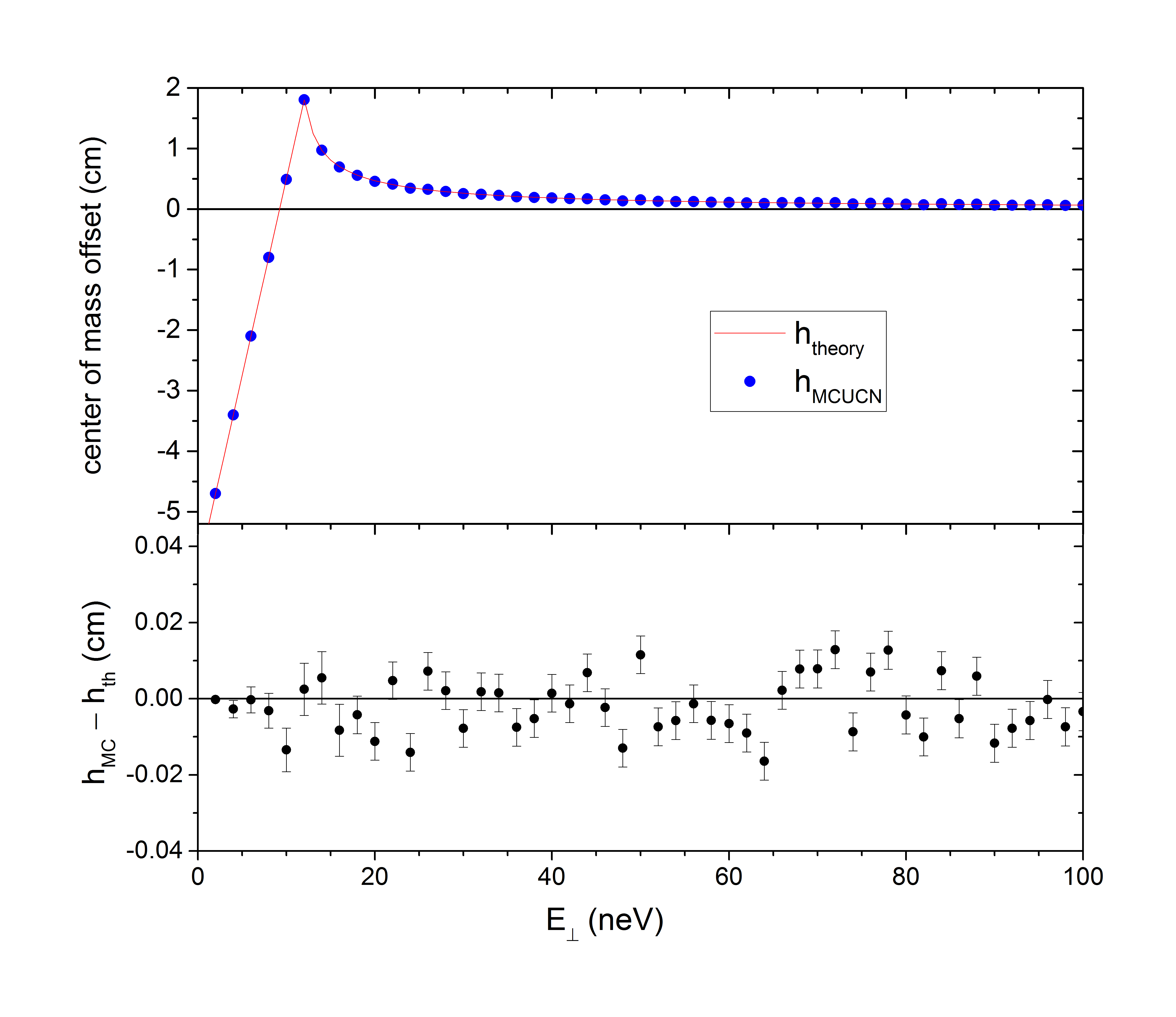}
}
\caption{Comparison of MCUCN and analytic calculations of the vertical center-of-mass offset (see text).} 
\label{fig:MCUCN-Analytic-GravityNoDiffuse}
\end{center}
\end{figure*}

The results are shown in Fig.\,\ref{fig:MCUCN-Analytic-GravityNoDiffuse} where the ensemble-averaged center-of-mass values (at a given 'snapshot' time) were plotted as a function of the vertical part of the kinetic energy. The residual between the simulation and the theoretical prediction can be seen in the lower graph. We could show that at the lowest energies (best statistics data points) we can approach the theory within 0.1\,\%, and at higher energies (lower statistics) the error bars are consistent with theory within $3\sigma$. This also demonstrates the very good accuracy of MCUCN in recording UCN coordinates at different 'snapshot' times $-$ a critical feature for the calculation of UCN densities.

\subsubsection{Diffuse and specular reflections}
\label{label-Diffuse-and-specular-reflections}

In a next test, we introduced diffuse reflections and compared theoretical and MC calculations of the center of mass of UCN. By including diffuse reflections, we generated conditions close to full mechanical equilibrium.

For the nEDM systematics it is important to calculate instead of the ensemble average the time average $\left\langle h \right\rangle _{time}$ of the center-of-mass offset. This is also a computing advantage because we get a higher precision by time-averaging as when averaging over a number of UCN. According to the ergodicity theorem, these two averages have to converge if a perfect mechanical equilibrium is reached. In this example the MC runs were performed at fixed total energy, i.e. the full kinetic energy plus the gravitational potential with reference to the bottom plane. The analytical approximation is given by eq. (5) in \cite{Harris2014}

\begin{equation}
\label{eq-aver-CoM}
    \left\langle h \right\rangle = - \frac{\epsilon}{k} \left[ 0.6 - k - 0.6 \left( 1 - k \right)^{5/3} \right],
\end{equation}

where $\epsilon=\frac{E_\text{tot}}{mg}$ is the kinetic energy at the bottom of the chamber in $mg$ units (we have to consider $E_\text{tot}$ and not only $E_\perp$ since the latter is not constant), $ k = 1 $ when $ \epsilon < H $, and $ k = 1 - \left( 1 - H/\epsilon  \right)^{3/2} $.

We varied the probability of diffuse reflections as input parameter over a large range from 0.1 $-$ 99.9\%. The angular distribution of generated UCN from the bottom was according to $P(\text{cos} \theta)d(\text{cos} \theta) = \text{const} \cdot \text{cos}\theta d(\text{cos}\theta)$ similar as in the previous calculation in subsection \ref{label-Specular-reflections}. 

\begin{figure*}[htb]
\begin{center}
\resizebox{0.80\textwidth}{!}{
\includegraphics{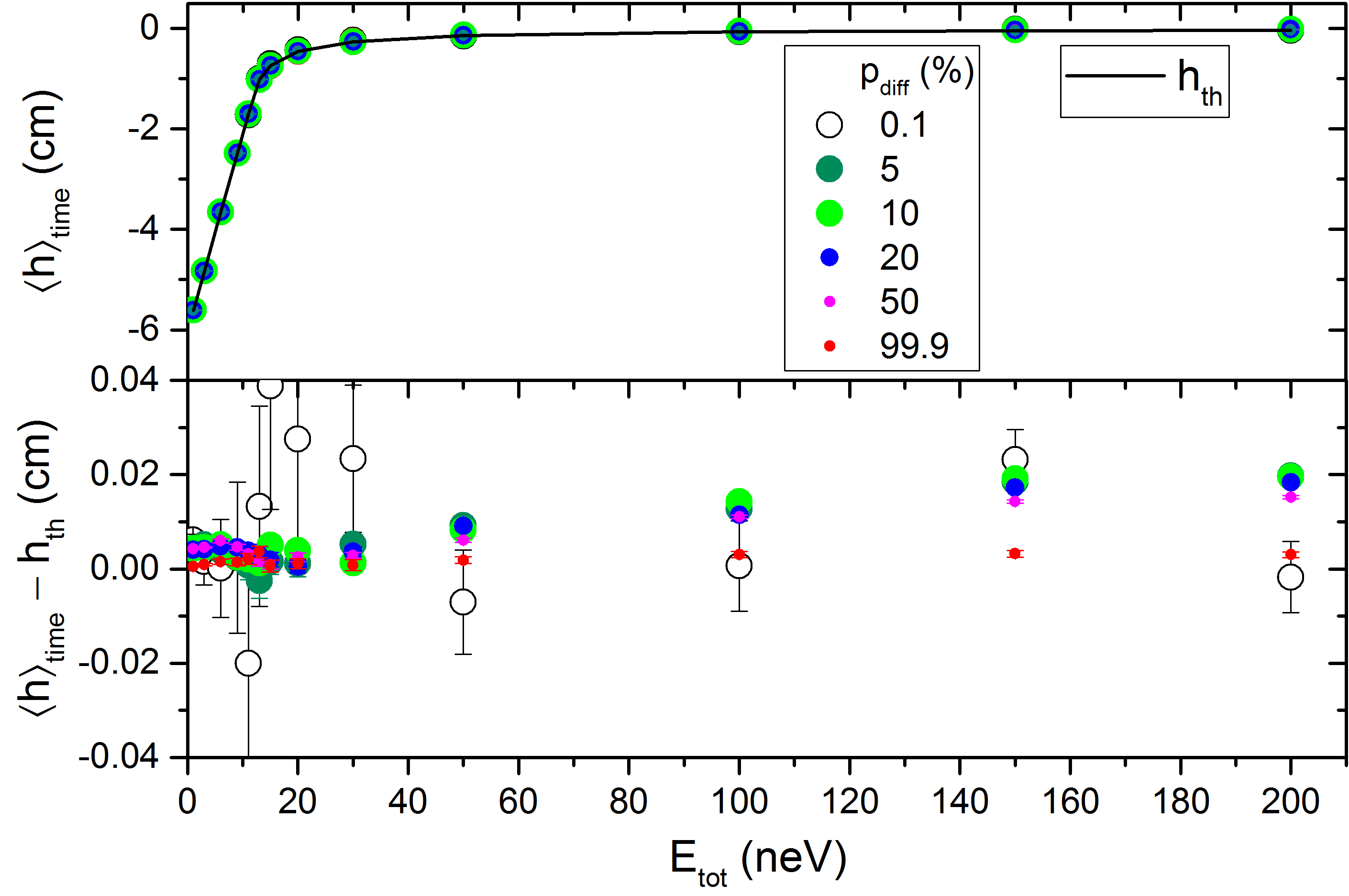}
}
\caption{Comparison of the MCUCN time-averaged center of mass offset $\left\langle h \right\rangle _{time}$ with the theoretical approximation $\left\langle h \right\rangle _{th}$ versus kinetic energy at the chamber bottom as a function of the fraction of diffuse reflections $p_\text{diff}$.} 
\label{fig:MCUCNcompare-CoMoffset-th-and-time-aver}
\end{center}
\end{figure*}

In Fig.\,\ref{fig:MCUCNcompare-CoMoffset-th-and-time-aver} we plotted the deviation of the simulated center of mass from the analytic prediction as a function of the total energy of the UCN. The analytic and the simulated values merge for larger diffuse reflections fraction. This was expected since the analytic approach assumes perfect mechanical equilibrium which was achieved faster if the fraction of diffuse reflections was set higher. 

We additionally carried out an ergodicity test, i.e. checking whether the time average $\left\langle h \right\rangle _{time}$ was equal to the ensemble average at a given time moment.  We could demonstrate that ergodicity was fulfilled with the available statistics within 0.2\%. For this test, a mechanical equilibrium was apparently achieved. Here as well, we concurrently tested the MCUCN algorithm of recording UCN coordinates at various 'snapshot' times.

\subsubsection{UCN loss rate}
\label{label-UCN-loss-rate}

The reliability of the algorithm controlling the UCN loss rate over time was tested. We compared the storage time constants obtained with MCUCN to analytic approximations as explained below. 
																		        
In the simulations we considered the same chamber setup as previously but switched on the loss parameter $\eta$. This is independent of the kinetic energy but depends on the optical potential of the walls. As in the real nEDM experiment, we set different coating materials for the cylinder surface (dPS with $V_\text{F}$ 165\,neV), and for the top and bottom surfaces (DLC with $V_\text{F}$ 220\,neV). For both surfaces the loss parameter $\eta$ was set $3\cdot 10^{-4}$, and the diffuse reflection fraction $p_\text{diff}$ to 0.50. An additional loss channel via beta decay with a time constant of 880\,s was included. 

In a first analytic approximation we considered as separate channels the losses on the side, top and bottom walls since these experience different bounce rates and have different optical potentials. For different walls we used the average kinetic energy i.e. the total energy minus the average potential energy. For example, in the simplest case, for the bottom wall this meant that the average bouncing kinetic energy was equal to the total energy at the bottom (potential energy zero). For the side wall the average kinetic energy is $E_\text{tot}-mg\left\langle h \right\rangle$ making use of eq. (\ref{eq-aver-CoM}). Although we considered gravity, for the average bounce rate estimation we relied on the approximation for an isotropic gas. This gives the average time between the bounces $4V /(S v)$, where $V$ is the volume, $S$ the total surface or a surface element for which the bounce rate is averaged, and $v$ the neutron velocity. The loss per bounce was calculated according to eq. (2.70) in \cite{Golub1991} performing isotropic angular averaging, which is also not exactly valid because of gravity.

The second analytic approximation is very similar except that instead of using a vertical average of the kinetic energy for calculating the bounce rate on the cylinder wall we used eq. (28) from \cite{Pendlebury1994} which takes more precisely into account the change in bounce rate along the vertical direction.

\begin{figure*}[htb]
\begin{center}
\resizebox{0.90\textwidth}{!}{
\includegraphics{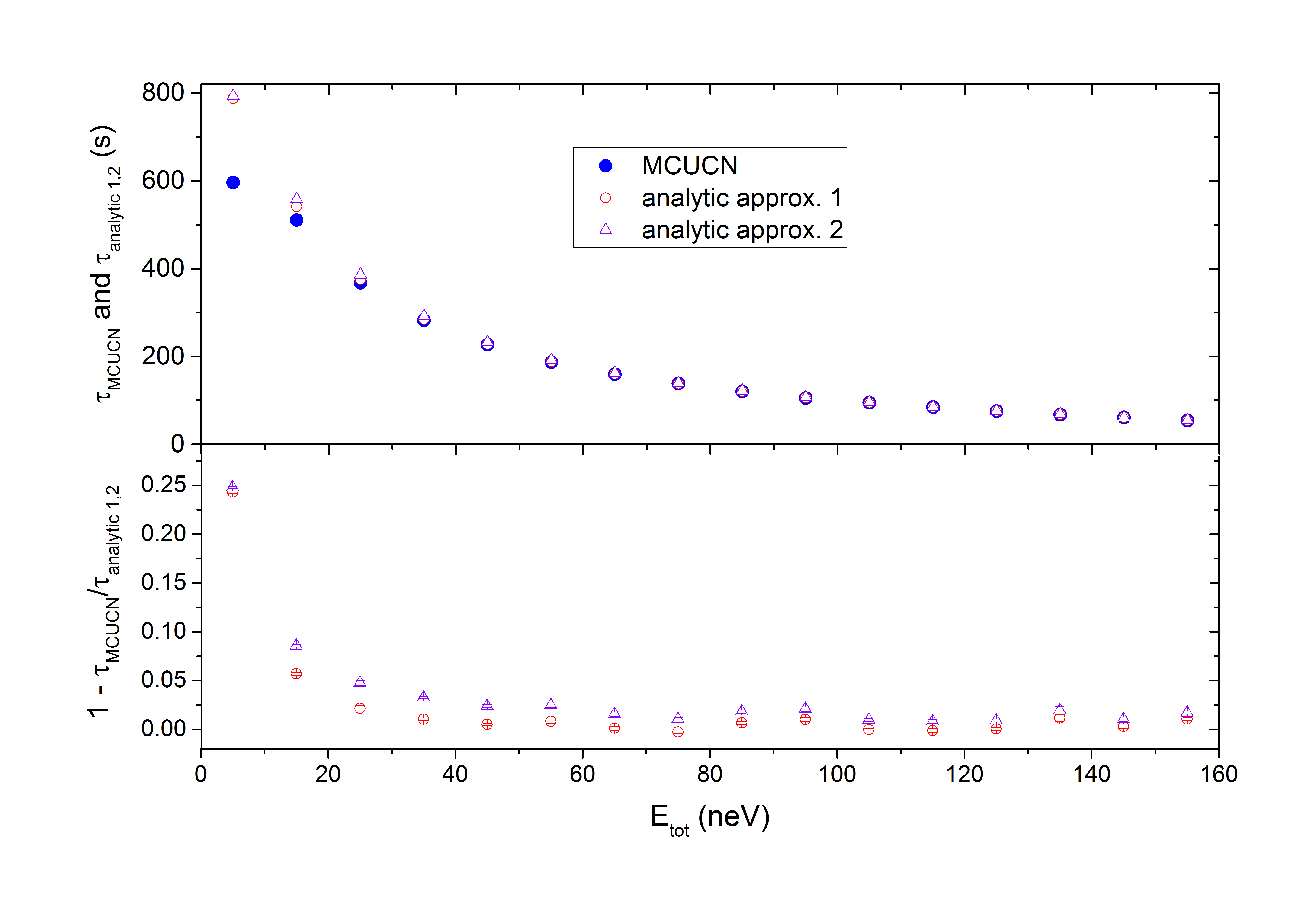}
}
\caption{Comparison of nEDM storage time constants as a function of energy, calculated with MCUCN and analytic approximations.} 
\label{fig:MCUCN-Analytic1-Analytic2-relativeDeviationsMCUCNanalytic1-2}
\end{center}
\end{figure*}

The results of the comparison can be seen in Fig.\,\ref{fig:MCUCN-Analytic1-Analytic2-relativeDeviationsMCUCNanalytic1-2}. The two analytic approximations almost coincide in the whole UCN energy range, however, differ from the MC simulation for the lowest UCN energies comparable to the gravitational potential energy. We attribute this deviation to the strong influence of anisotropy due to gravity in this energy interval. 

The first analytic approximation is closer to the simulated values, as shown in the bottom of Fig.\,\ref{fig:MCUCN-Analytic1-Analytic2-relativeDeviationsMCUCNanalytic1-2}. This means that the simplification done in the first approximation, i.e. the aforementioned energy-averaging, happens to slightly compensate for neglecting the anisotropy. The relative deviation between MCUCN values and first analytic approximation above 35\,neV is spread around 0.5\,\% with SD 0.5\,\% proving the correct implementation of the UCN losses. 

As a technical detail, the computing speed of one MCUCN data point was 15\,ms per trajectory on an AMD Opteron(TM) 6282 SE processor.

\subsection{Spin precession}
\label{label-benchmarks-spin}

In this section we demonstrate the capabilities and accuracy of the spin-handling algorithm in MCUCN\@. In the first simple example the simulation of a spin flip following a uniformly rotating magnetic field was compared to the exact analytic solution. In further examples we also introduced a magnetic field inhomogeneity and studied its effect on the Ramsey resonance frequency. In the last example of this section we compared MCUCN results with an analytic calculation of a false nEDM signal originating from a relativistic effect experienced by UCN moving in an electric field parallel to the inhomogeneous magnetic field. In order to separate transverse depolarization, the spin-flip probability at surface collisions was always set to zero.

\subsubsection{Spin-flip in a rotating magnetic field - adiabaticity check}
\label{label-Spin-flip-rotating-field}

In this first example we considered a magnetic-field vector of 1\,G and a neutron with a polarization vector parallel to this field $-$ both parallel to Y $-$ at initial time $t =$ 0\,s. The field vector rotated by $\pi/2$ around the Z axis with final direction parallel to X with an angular speed which could vary. Under these conditions, the UCN polarization vector will adjust to the new field direction depending on the ratio of Larmor frequencies of the UCN and the field-vector rotation frequency. This latter ratio is called adiabaticity \cite{Golub1991}.

In the extreme case when the adiabaticity parameter was very large, the simulated UCN polarization followed perfectly the magnetic-field direction. If it was very small the polarization vector did not change direction at all in the laboratory system. This problem has an exact analytic solution for intermediate adiabaticity cases as described in \cite{Newton1948}, and was already used as a benchmark earlier in \cite{Seeger2001,G4UCN}.

\begin{figure*}[htb]
\begin{center}
\resizebox{0.90\textwidth}{!}{
\includegraphics{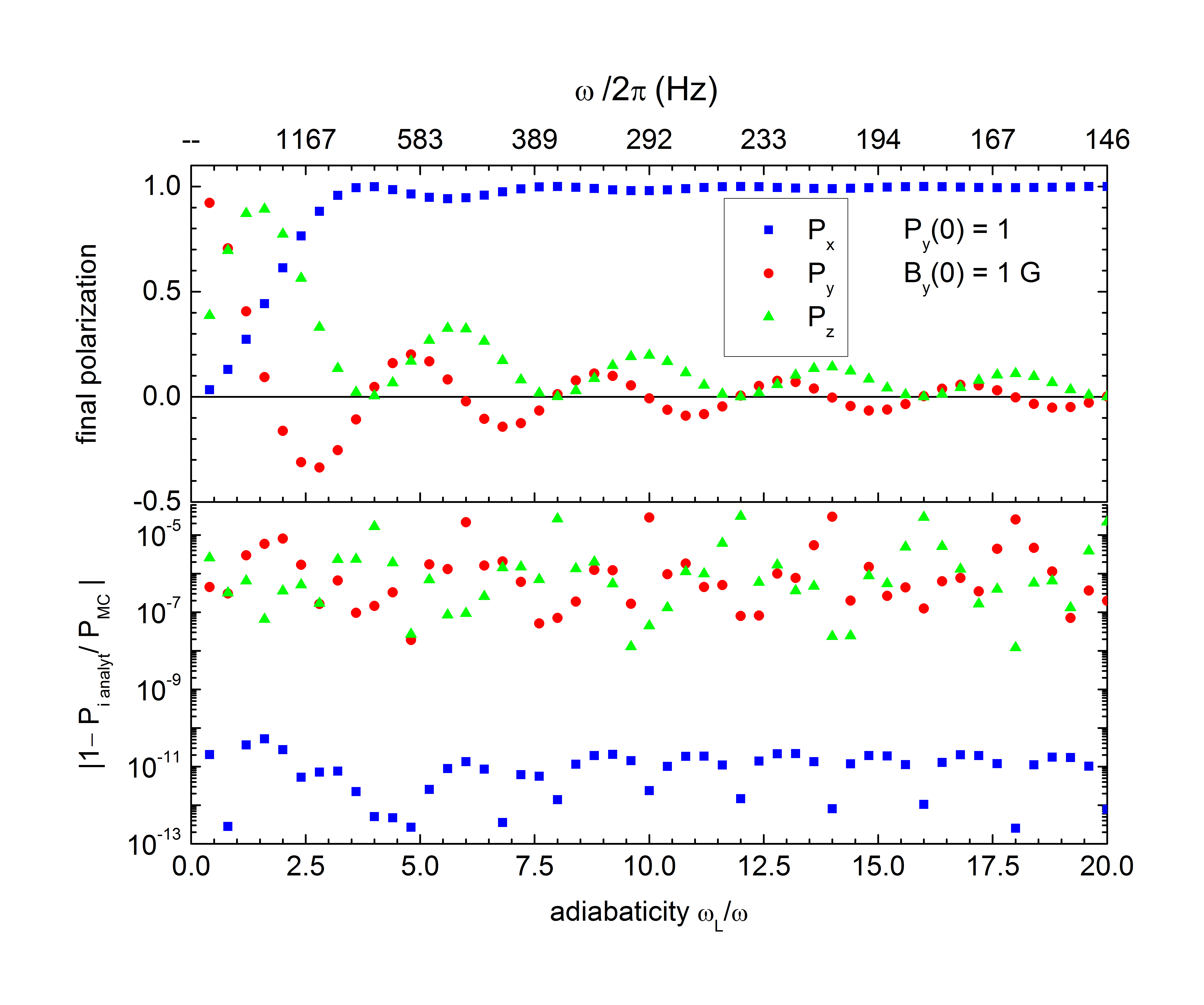}
}
\caption{Comparison of results from the MCUCN spin handling algorithm with the exact analytic solution. We calculate the final polarization after $\pi/2$ field rotation as a function of the adiabaticity coefficient.} 
\label{fig:MCUCN-90degRotation-analytic-vs-MC}
\end{center}
\end{figure*}

The comparison of MCUCN simulation results and the analytic solution is plotted in Fig.\,\ref{fig:MCUCN-90degRotation-analytic-vs-MC}. The top part shows the final polarization projection components $P_{x,y,z}$ after the field vector completely rotated $\pi/2$ as a function of the adiabaticity parameter. The bottom plot shows the relative deviations from the analytic solution for all three polarization projections. The precision obtained was better than $3 \cdot 10^{-5}$. In the very top scale the rotation frequency of the field vector was indicated. For larger adiabaticity, i.e. lower field rotation frequency, the relative error of the large final component, $P_x$ is smaller than $10^{-10}$, however, this extremely high accuracy was not achieved in the low-adiabaticity range by the large components $P_{x,y}$. This is because the field rotation frequency here is large and relatively less well handled by the spin handling algorithm compared to lower frequencies $-$ however, it is still accurate enough for our purposes.

As a technical detail, the average time consumed by the algorithm with the same processor to obtain one data point at this precision was 0.18\,s.

\subsubsection{Full Ramsey pattern}
\label{label-Ramsey-pattern}

\begin{figure*}[htb]
\begin{center}
\resizebox{0.90\textwidth}{!}{
\includegraphics{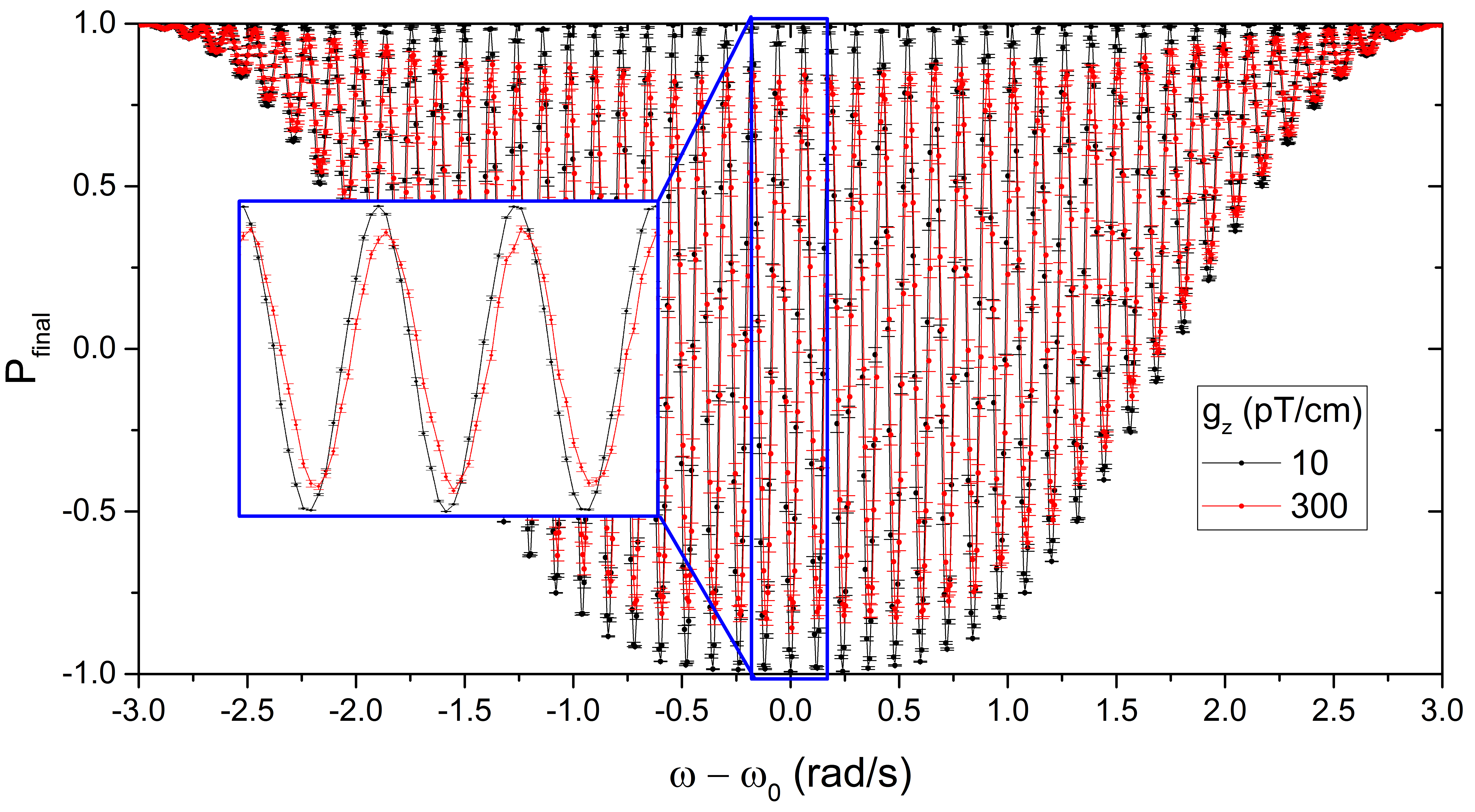}
}
\caption{MCUCN simulation reproducing full Ramsey scans with vertical gradients demonstrating the gravitational frequency shift.} 
\label{fig:MCUCN-RamseyScanWitVertGradients}
\end{center}
\end{figure*}

In all previous searches for an nEDM the Ramsey method of separated oscillatory fields was used \cite{Harris1999}. Full Ramsey scans could be well reproduced with MCUCN, including the exact modeling of the $\pi/2$ flips, as shown in Fig.\,\ref{fig:MCUCN-RamseyScanWitVertGradients}. Here we performed the calculations with two vertical field-gradient values: 10 and 300\,pT/cm (compare to Fig.2 in \cite{Harris1999}. The insert in this figure shows a zoom into the central fringe region displaying a sinusoidal shape. In previous simulations \cite{Bodek2011}, we already demonstrated that a theoretical fit of this central region  reproduced well (i) the fringe pattern period $1/(T + 4t_{\pi/2}/\pi)$, including the phase parameter $4t_{\pi/2}/\pi$ characteristic for the Ramsey cycle, where $t_{\pi/2}$ is the duration of the $\pi/2$ pulse; (ii) the predicted frequency shift caused by the gradient and the center of mass offset; (iii) for one UCN energy the quadratic dependency of the transverse depolarization time-constant on the applied field gradient. 

Starting from these simple test cases, it will be possible to calculate frequency shifts and depolarization effects due to higher order gradients and/or a combination of these. For differing shift values for the parallel and anti-parallel magnetic and electric field configurations the size of the false nEDM can then be estimated.

\subsubsection{Bloch-Siegert-Ramsey shift}
\label{label-Bloch-Siegert-Ramsey-shift}

\begin{figure*}[htb]
\begin{center}
\resizebox{0.90\textwidth}{!}{
\includegraphics{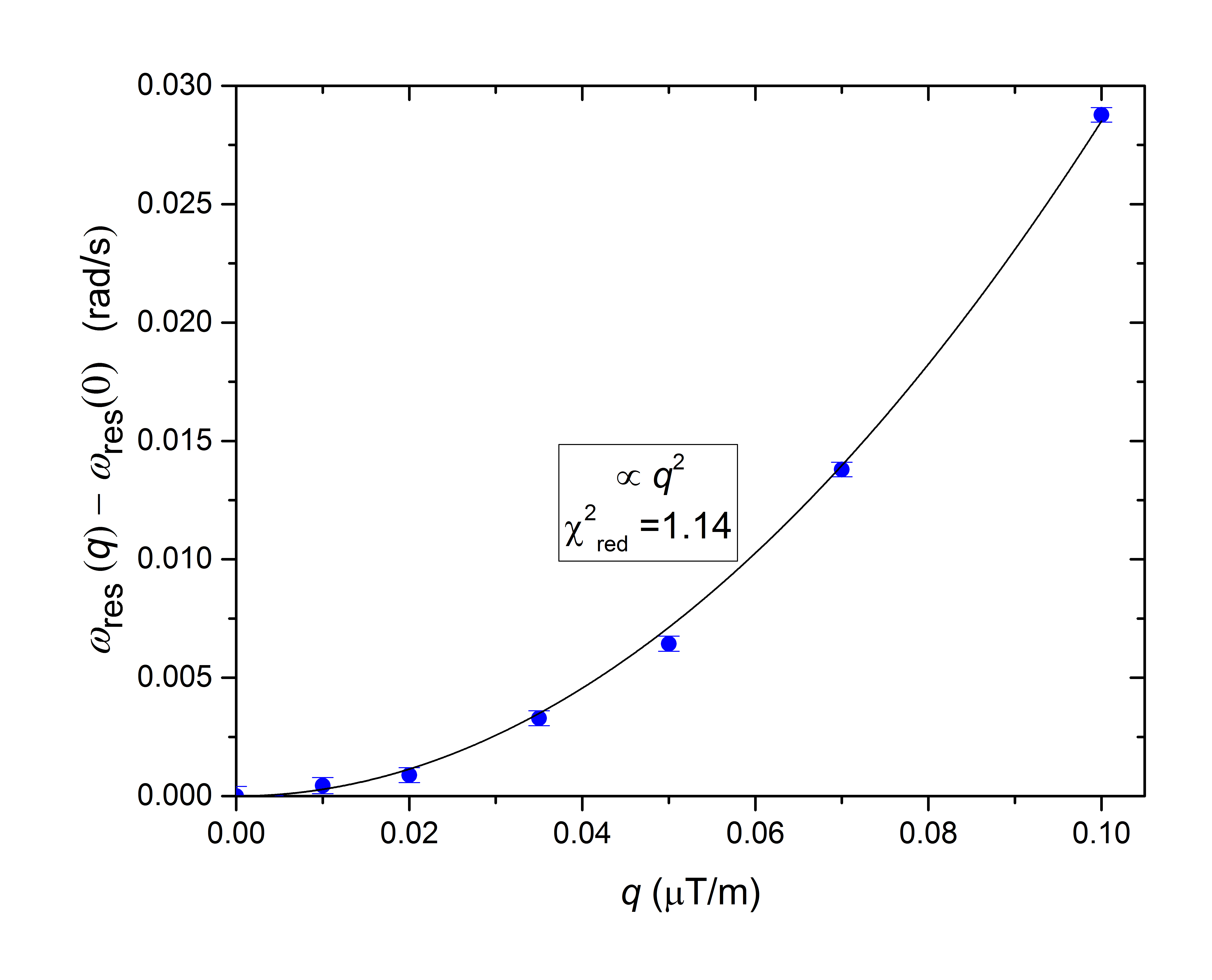}
}
\caption{Benchmark of MCUCN simulations by comparing to the analytical formula for the Bloch-Siegert-Ramsey shift, see text.} 
\label{fig:MCUCN-BlochSiegertRamseyShift-Benchmark}
\end{center}
\end{figure*}

One source of systematic effects in the nEDM measurement, the Bloch-Siegert-Ramsey shift, was first described in \cite{Pendlebury2004} using as a starting point the work of Ramsey \cite{Ramsey1955}. The subject was treated recently in \cite{Pignol2012}.  According to these studies, a resonance frequency shift is induced by a rotating transverse magnetic field component (seen in the frame moving with the UCN), which can be for example a quadrupole field. If in an nEDM measurement the quadrupole field strength happens to be different for the parallel and the anti-parallel magnetic and electric field configurations, this will generate a false signal. In the aforementioned studies it was shown that the resonance frequency shift goes quadratically with the quadrupole strength parameter $q$ defined in the vector expression  $(B_x\ \ B_y) = q(y\ \ x)$. This could be very well reproduced by MCUCN as shown in Fig.\,\ref{fig:MCUCN-BlochSiegertRamseyShift-Benchmark} where simulation and theory agree within $2\sigma$.

\subsubsection{False nEDM from the motional magnetic field}
\label{label-False-nEDM-motional-field}

As proven in \cite{Pendlebury2004}, one systematic effect in the nEDM measurement is the false signal coming from a combination of transverse field gradients and the motional magnetic field seen by the UCN in the presence of an electric field (\textbf{E}) $\textbf{B}_v = (\textbf{E}\times \textbf{v})/c^2$, where v is the velocity vector of the UCN. The electric field is parallel or anti-parallel to the main $B_0$ field.

\begin{figure*}[htb]
\begin{center}
\resizebox{0.80\textwidth}{!}{
\includegraphics{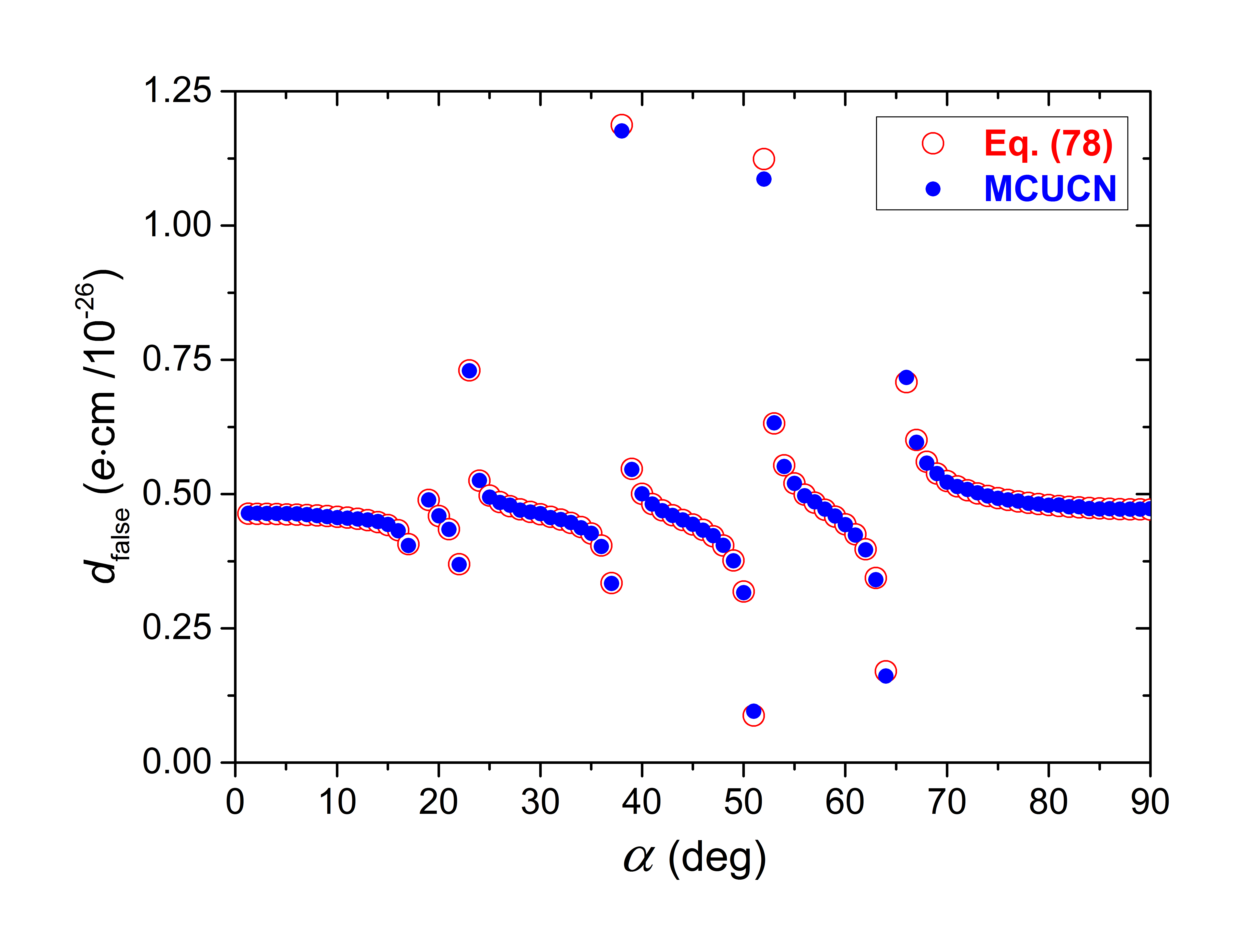}
}
\caption{Analytic benchmark of spin handling: MCUCN simulation reproducing the false nEDM calculation in \cite{Pendlebury2004}.} 
\label{fig:MCUCN-Eq-78-PendleburyPhysRevA70-2004-032102}
\end{center}
\end{figure*}

In this simulation test example considering a simplified case, we reproduced the theoretically predicted false nEDM values given by eq.(78) in \cite{Pendlebury2004} as a function of the radial angle difference ($\alpha$) between two reflection points of orbiting UCN along the side-wall of the cylinder chamber, as depicted in Fig. 3 of \cite{Pendlebury2004}. The horizontal velocity was set constant to 5\,m/s and diffuse reflections were set to zero. The uniform vertical magnetic-field gradient was set to $\partial _z B_z$ = 1\,nT/m which along with the cylindrical symmetry results in a radial field $\textbf{B}_r = -\partial _z B_z \textbf{r} /2$, i.e. linear in radius. In Fig.\,\ref{fig:MCUCN-Eq-78-PendleburyPhysRevA70-2004-032102} we compare MCUCN and analytic results demonstrating excellent agreement.

\section{Application examples}
\label{label-applications}

The MCUCN code was extensively used in testing our nEDM data analysis methods including studies of gravitational depolarization and systematic effects \cite{Pendlebury2015,Afach2015PRL,Afach2015PRD}. Further, we used the code also for supporting the analysis of measurements for the comparison of ultracold neutron sources worldwide \cite{Bison2017}.

Below, we briefly explain several recent applications of MCUCN also including simulation results not detailed in the aforementioned publications, however, important to demonstrate the performance of the code.

\subsection{Full simulation model of the PSI UCN source and beamlines}
\label{label-mcucn_model_psi_source_beamlines}

A first main focus in building the MCUCN code was the development and characterization of the guiding and storage system of the PSI UCN source. A full description of the UCN source and beamline guides as built was provided in \cite{Ries2016,Goeltl2012,Lauss2012,Blau2016}.

In the MCUCN model of the PSI UCN source and beamlines we included all details of surfaces interacting with UCN having characteristic dimensions above 10\,cm. We also added important gaps which were set as either completely absorbing or with a low nuclear optical potential like that of aluminum. 

Here we summarize geometric details implemented in the model: An aluminum lid above the sD$_2$ converter, approached by a spherically curved surface, constitutes the lower boundary of the vertical cylinder-shaped guide which directs the UCN into a large 2 m$^3$ storage volume. This latter volume has four flat sides. In the bottom of one of these, a niche for the South and West-1 beamlines, and in the top a niche for beamline West-2 were constructed. Two half-disk flaps with time-dependent opening angles separate the storage volume from the vertical guide sitting below it. Disk flaps can separate this source volume from the beamlines. All three beamlines have a 30 deg kink. The beamline guide sections \cite{Blau2016} correspond to the engineering drawings within 2 mm precision. Vacuum separation foils made from 0.1 mm AlMg3 constitute the end of the three beamlines. On beamline South the foil was placed in the center of a 5 T superconducting magnet working as UCN polarizer. This was implemented in MCUCN as a spin-dependent rectangular potential barrier (for spin-up) or well (for spin-down) located at the foil. A test volume was defined at the end of the West-1 beamline by two disk shutters which move either continuously or switched on and off. Virtual detectors with ideal efficiency count the UCN passing through the 0.1 mm thick detector separation foil. All surface sections, for which at least one surface parameter could be different from the neighboring one, were implemented separately. At the end of the South beamline we constructed the UCN guide and storage system of the current nEDM apparatus and, as alternative, an upgrade to a double-chamber (n2EDM) apparatus.

The nuclear optical-potential values of the coatings were set as obtained from previous cold neutron reflectometry measurements, 220 neV for DLC \cite{Grinten1999,Atchison2007b} and similar for NiMo \cite{Goeltl2012}. Further parameters were determined from a number of test experiments with UCN \cite{Ries2016,Goeltl2012} which all could be well reproduced by MCUCN after we calibrated the unknown parameters. This specific calibration of the MCUCN model parameters will be discussed in upcoming publications \cite{Blau2017SourceBeamlines,Blau2017Pingpong}. 

\begin{figure*}[htb]
\begin{center}
\resizebox{0.99\textwidth}{!}{
\includegraphics{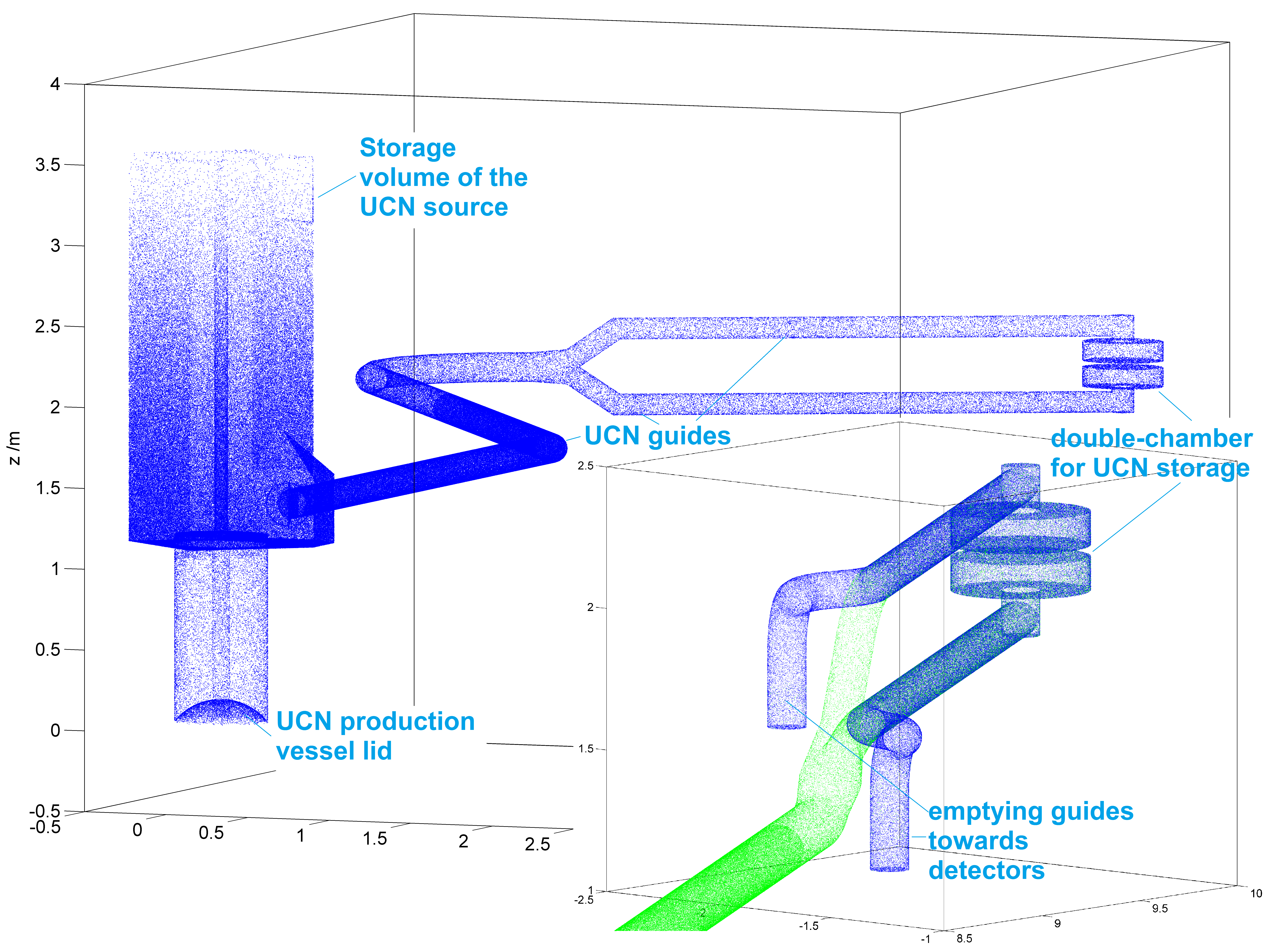}
}
\caption{Reflection points test visualizing the UCN source and beamlines at PSI together with the double-chamber setup for the n2EDM experiment.} 
\label{fig:MCUCN-UCNsourceANDn2edmLabels}
\end{center}
\end{figure*}

In the simulation of experiments it is important to consider the whole geometry of the filling process because UCN travel multiple times along the guides, repeatedly entering and exiting the source volume and the experimental volume. In other situations, for example emptying the experiment chambers, a separate simulation can be performed using as input the results of a previous simulation. Both cases are illustrated in Fig.\,\ref{fig:MCUCN-UCNsourceANDn2edmLabels}.

In Fig.\,\ref{fig:MCUCN-UCNsourceANDn2edmLabels} we display the UCN source geometry along with the n2EDM double-chamber experiment by plotting the calculated reflection points (blue dots) also as a test for the algorithms. The smaller insert shows the reflection point test of the emptying geometry of the n2EDM double chambers (blue dots - emptying phase, and green dots the previous filling phase).

\subsubsection{Simultaneous detection efficiency estimation for a possible n2EDM setup}
\label{geometry-optimization-experiment}

The current nEDM experiment at PSI makes use of a simultaneous detection system for both spin components (USSA) as detailed in \cite{Afach2015USSA}. This type of setup will be also used in the next generation n2EDM experiment. Furthermore, n2EDM will have two precession chambers at different elevations, and thus will store slightly different energy spectra. 

\begin{figure*}[htb]
\begin{center}
\resizebox{0.99\textwidth}{!}{
\includegraphics{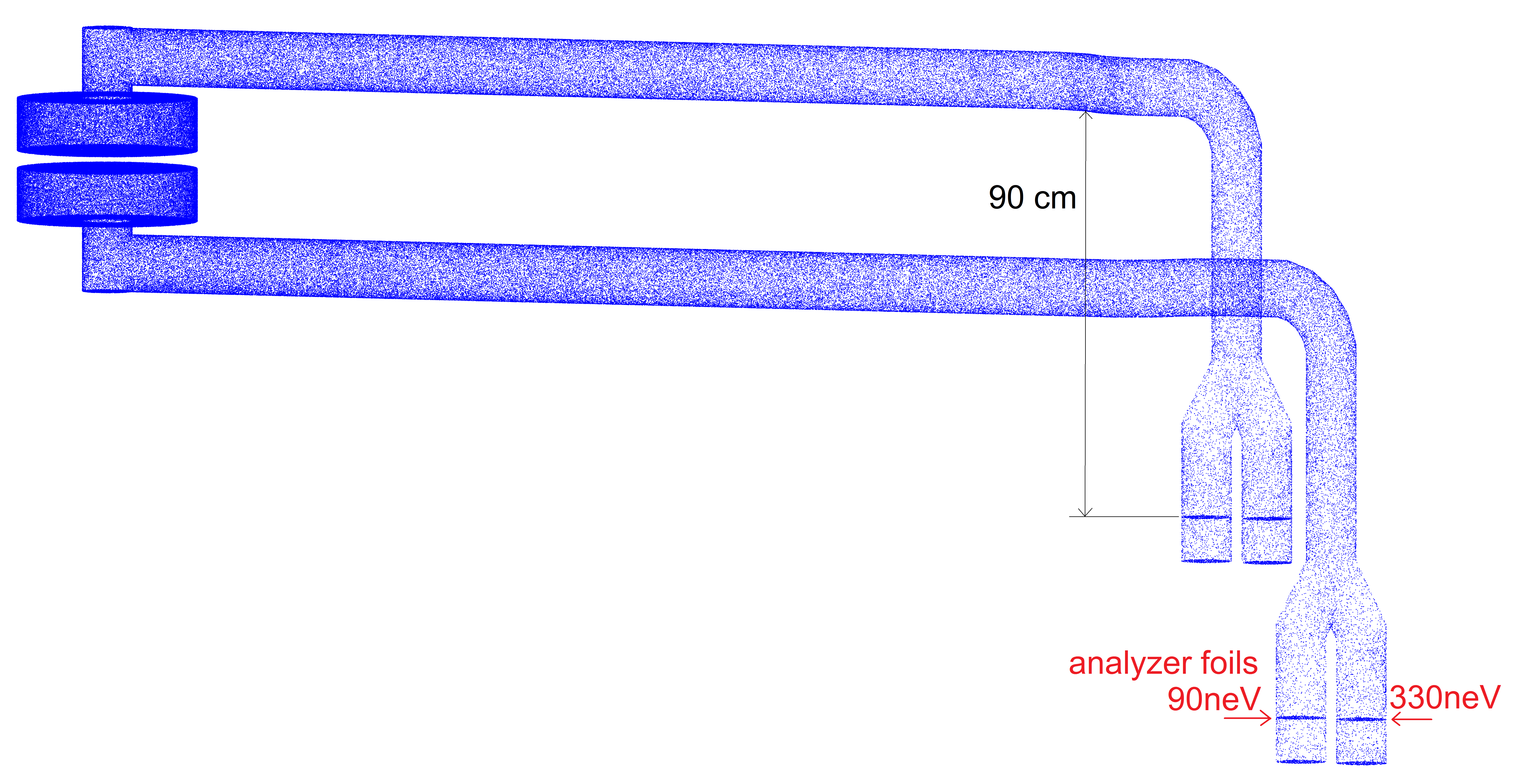}
}
\caption{Reflection points visualization of the UCN emptying and detection for a possible n2EDM setup. The analyzer foil positions and the fall-height are indicated along with the optical-potentials experienced by the two spin states.} 
\label{fig:MCUCN-n2EDMfullDetection}
\end{center}
\end{figure*}

In Fig.\,\ref{fig:MCUCN-n2EDMfullDetection} we show a plot of the geometry as being strobed by the UCN (blue = reflection points) during the phase of emptying toward the detectors. The two chambers modeled here have independent simultaneous spin analyzer systems both consisting of two arms. 

In the real system, the UCN are guided to a magnetized analyzer foil which acts as an energy filter as a function of the spin state. A UCN with spin anti-parallel to the holding magnetic-field experiences a 90\,neV potential barrier, while the spin parallel to this field sees a higher 330\,neV barrier \cite{Golub1991} and is reflected since the arriving energy spectrum should be fully located below this threshold value. On the second arm of the USSA an adiabatic spin flipper (ASF)\cite{Geltenbort2009} is switched on, thus allowing for the detection of the UCN with spin state reflected from the first USSA arm, or the same spin state coming directly from the chamber.

In order to transmit all UCN a 90\,cm falling height is used to shift the UCN energy spectrum to start at approximately 90\,neV.

The efficiency of a real ASF can be routinely adjusted to 99.9\%. Thus in the simulation model we can simplify the calculation of emptying efficiency by setting in one arm the optical potential of the analyzer foil to 90\,neV, and in the other arm setting 330\,neV. 

Using this model we can compute the detection efficiency of the UCN during emptying the n2EDM chambers. We can do this as a function of the total energy defined relative to the bottom of each chamber, i.e. the kinetic energy at the bottom. We assumed that the same optimized detection system will be later applied for both upper and lower n2EDM chambers.

In the example presented here we set the guide loss parameters $\eta$ to $ 3\cdot 10^{-4}$ and include several gaps of 0.2\,mm located in the kinks of guide sections. The coating parameters of the precession chambers were the following: both insulator (cylinder side) and electrodes (top and bottom) had an optical potential 220\,neV and a uniform loss parameter $3\cdot 10^{-4}$. One ASF was 'on' meaning that we set different optical potential for the analyzer foils as described above.

\begin{figure*}[htb]
\begin{center}
\resizebox{0.70\textwidth}{!}{
\includegraphics{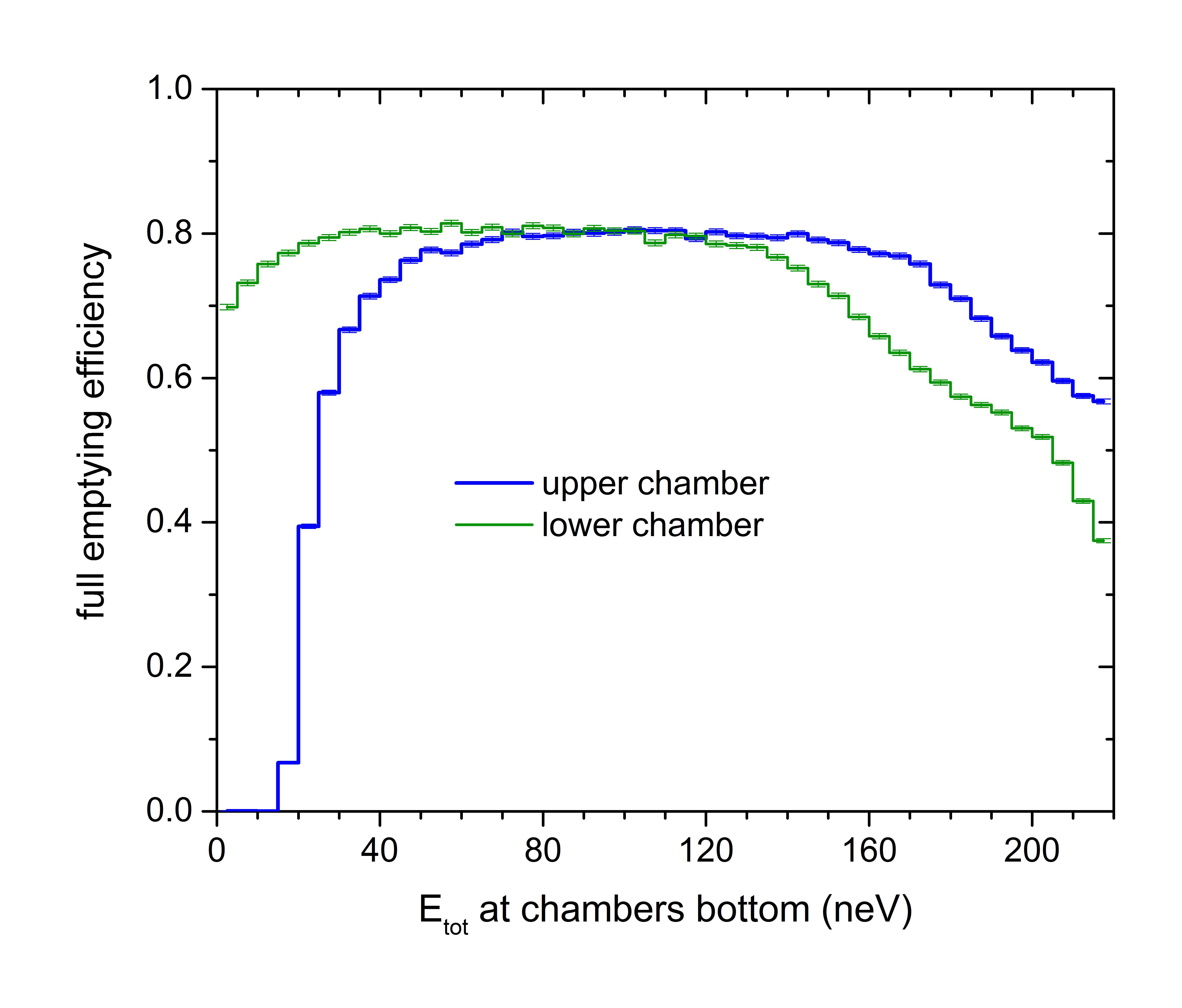}
}
\caption{The detection efficiency for the top and bottom chambers in the n2EDM setup as a function of the kinetic energy at the bottom of the respective chambers.} 
\label{fig:MCUCN-n2EDMfullDetectionEfficiency}
\end{center}
\end{figure*}

The calculated detection i.e. emptying efficiency for both chambers was plotted in Fig.\,\ref{fig:MCUCN-n2EDMfullDetectionEfficiency} as a function of kinetic energy at the bottom of each corresponding chamber. There is a difference between the upper and lower chamber results. This is caused by the different positions, top or bottom,  of the emptying/filling openings. From separate calculations of the energy spectra of the UCN after storage in both chambers we estimated an overall emptying efficiency of about 70\% in UCN counts.

\subsection{nEDM experiment systematics}

\label{label-applications-nEDM}

One other important application of MCUCN was supporting nEDM systematics calculations and related effects. Here we show examples in which the physics part related to gravitational resonance frequency shift and depolarization was treated in detail in \cite{Pendlebury2015,Afach2015PRD}. 

For details of the model we refer to the benchmark examples discussed in Sec. 3. 
 and references therein.

\subsubsection{Linearity check of the resonance curve as a function of vertical gradient}
\label{label-Linearity-check-resonance-curve}

\begin{figure*}[htb]
\begin{center}
\resizebox{0.70\textwidth}{!}{
\includegraphics{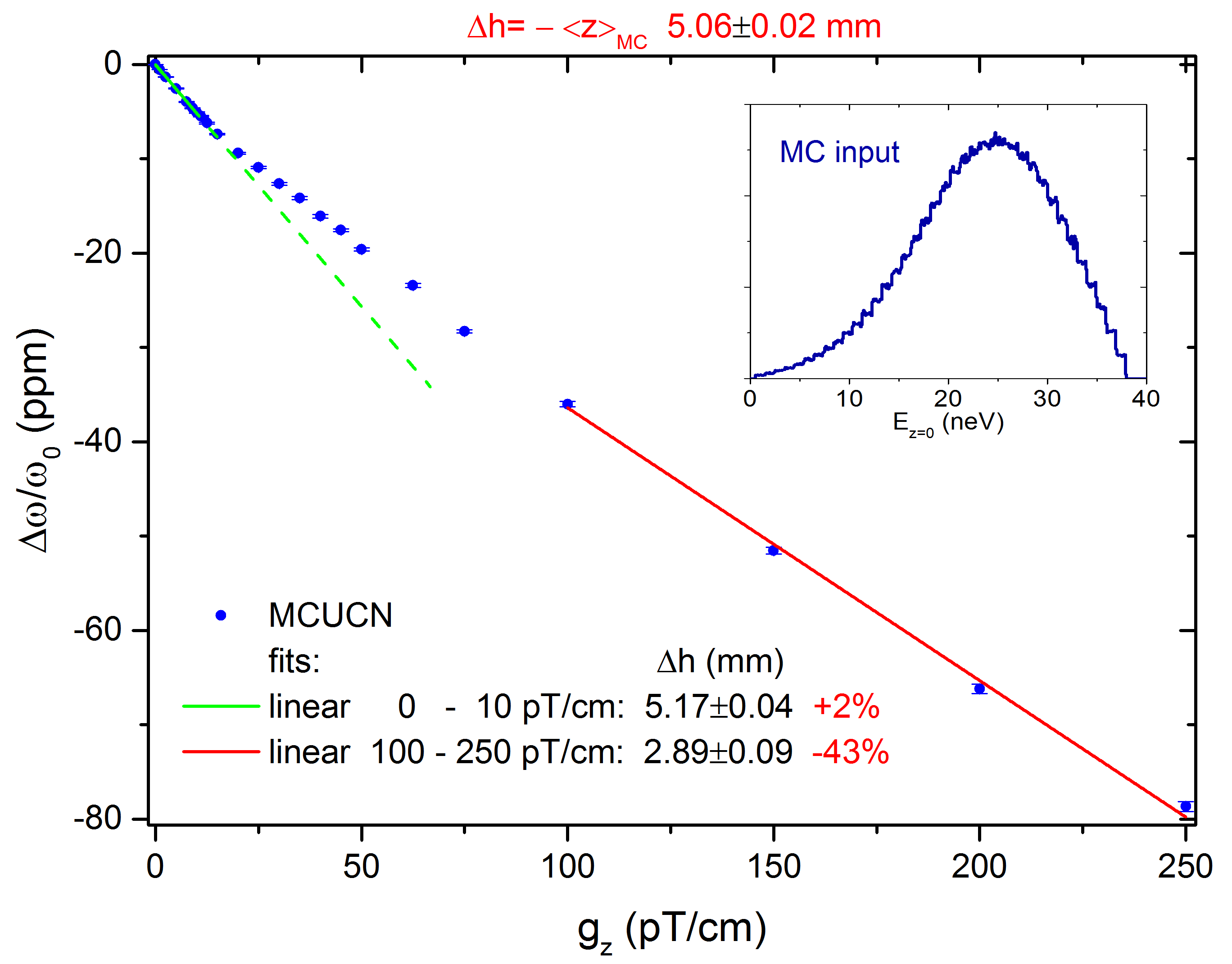}
}
\caption{Simulated relative frequency shift (in ppm) due to gravitational depolarization as a function of the linear magnetic field gradient $g_z$. The insert indicates the input spectrum. The labels include the linear fitting ranges yielding different apparent center-of-mass offsets. The correct center-of-mass offset is obtained only in the lowest gradient range fit, matching the simulated average vertical position (see text).} 
\label{fig:MCUCN-R-curve-trumpet-g_z-fit-vs-mc-offset-5p1mm-gravitational-depolarization}
\end{center}
\end{figure*}

As first described in \cite{Baker2006}, the nEDM signal has to be modulated with the gravitational shift coming from a vertical linear magnetic-field gradient in order to interpolate to zero gradient. This can be achieved by scanning this gradient for both directions of the applied main magnetic-field vector (parallel and anti-parallel to gravity). Back then, a linear interpolation method was used by the expression $\Delta \omega / \omega_0 = h/B_0\ g_z$, where $h$ denotes the center of mass offset from the cylinder center and $g_z=\partial _z B_z$ a constant gradient. Later, in \cite{Harris2014} it was demonstrated theoretically and using simulations that the linear approximation used earlier will not work for larger gradients. This is caused by a 'wrap-around' effect meaning that a large dephasing exceeding $2\pi$ can show up as small dephasing. This will also diminish the rate of a frequency shift and depolarization (see also next subsection). In our example displayed in Fig.\,\ref{fig:MCUCN-R-curve-trumpet-g_z-fit-vs-mc-offset-5p1mm-gravitational-depolarization} we recalculated this nonlinearity effect using an energy spectrum (see insert) close to what we expected in the nEDM experiment \cite{Afach2015PRL}. In this figure we also compare the extracted center-of-mass offset of UCN as a function of example fit intervals. As expected, the true center-of-mass offset which value is indicated above the top axis can only be well reproduced when fitting in a range close to the origin. A fit performed at larger gradients underestimates the center of mass offset by 40\,\%. These studies led us to perform the magnetic field gradient scans in the experiment at lower values, in a range of $\pm$25\,pT/cm.

\subsubsection{Depolarization profile as a function of vertical gradient}
\label{label-Depolarization-profile-vertical-gradient}

\begin{figure*}[htb]
\begin{center}
\resizebox{0.80\textwidth}{!}{
\includegraphics{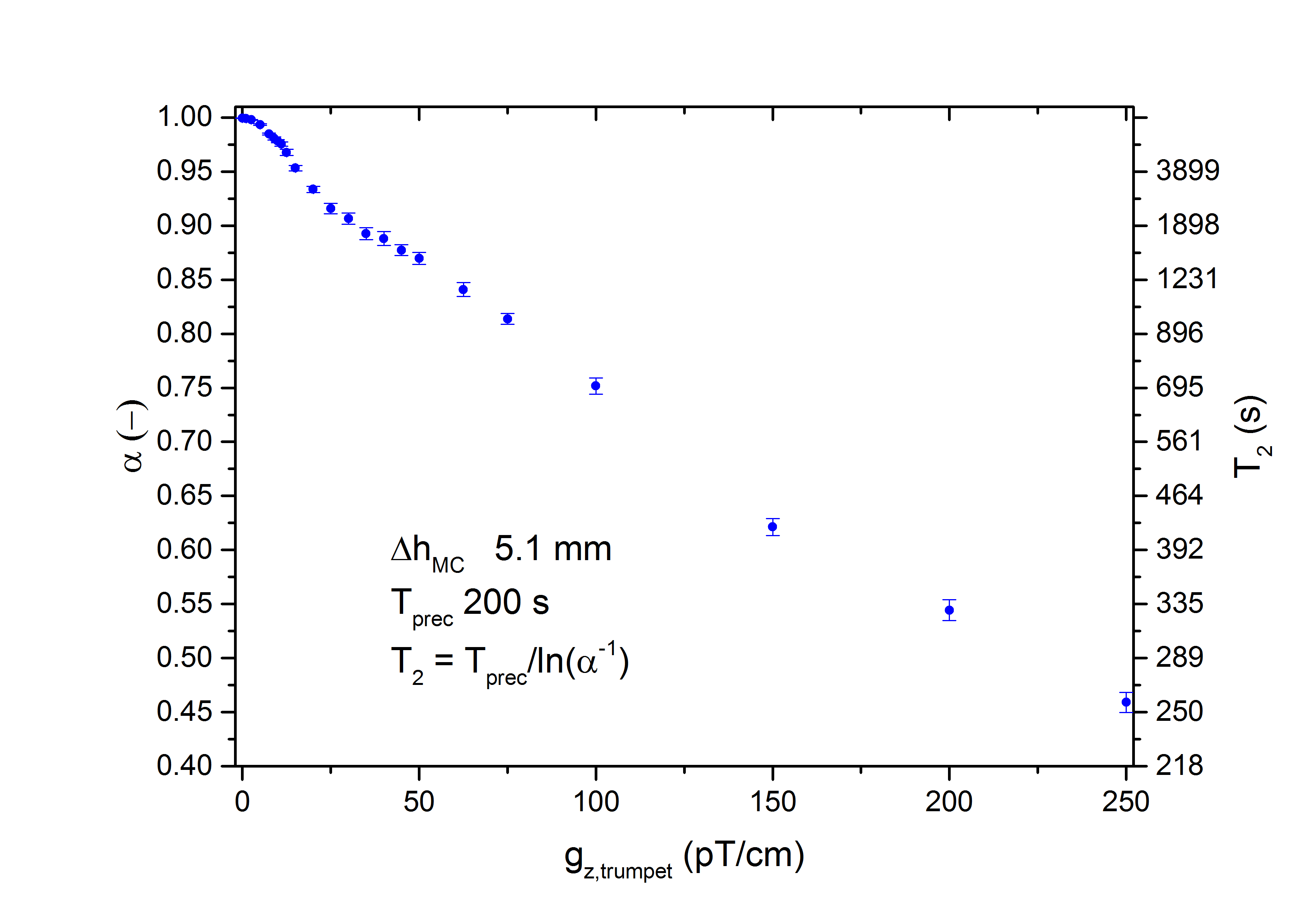}
}
\caption{MCUCN simulation of gravitational depolarization revealing the 'wrap-around' effect resulting in a reduced steepness of the slope at higher gradients above 20 pT/cm.} 
\label{fig:MCUCN-visibility-enhancedT2-v-gz-trumpet-5p1mm-UCNSEspectrum-morepoints}
\end{center}
\end{figure*}

The very same 'wrap-around' effect briefly explained in the previous subsection \ref{label-Linearity-check-resonance-curve} is visible in the depolarization profile taken as a function of the applied vertical magnetic-field gradient.  At higher gradients the slope of the curve becomes smaller \cite{Harris2014}. MCUCN was used to calculate this effect as plotted in Fig.\,\ref{fig:MCUCN-visibility-enhancedT2-v-gz-trumpet-5p1mm-UCNSEspectrum-morepoints}. 

MCUCN was also used for providing UCN energy spectra for the theoretical estimations which were compared to experimental data (at storage times typically applied in nEDM measurements) \cite{Afach2015PRD}.

\section{Conclusions}
\label{label-conclusions}

By comparisons to analytic models we demonstrated that the MCUCN code is a powerful and reliable tool for simulations of UCN physics experiments. 

Its applications so far extend from optimizations of experiments to estimations of nEDM systematics, e.g. gravitationally induced shifts of the Ramsey frequency and depolarization $-$ reflected by recent and upcoming publications \cite{Pendlebury2015,Afach2015PRL,Afach2015PRD}. 

Inter-comparisons with STARucn \cite{STARucn} simulations have also been very successful and will be reported in a future publication.

Decoupling visualization from numerical calculations makes MCUCN to a transparent and easily extendable tool also for non-expert programmers, and flexible for grid computing. The portability of the executable relying only on standard libraries makes it also to an ideal tool for computations on grids. 

MCUCN is made available for users who contact us, thus making sure that appropriate support can be provided.

\clearpage

\section*{Acknowledgements}

This work benefited from numerous discussions with M. Daum, B. Lauss, K. Kirch, P. Schmidt-Wellenburg (PSI Switzerland), and G. Pignol (LPSC France). We also greatly acknowledge granting access to the computing grid infrastructure PL-Grid \cite{PLGrid}.


\bibliography{MCUCN-references}

\end{document}